\documentclass[a4paper,11pt]{article}
\usepackage{pos}
\usepackage{todonotes}
\usepackage{siunitx}
\usepackage{setspace}
\usepackage{subfig}
\usepackage{wrapfig}
\usepackage{hyperref}

\title{The  Wavelength-shifting Optical Module for the IceCube Upgrade}
\ShortTitle{IceCube Upgrade WOM}

\author{The IceCube Collaboration \\{\normalsize \normalfont(a complete list of authors can be found at the end of the proceedings)}}

\emailAdd{jorackhe@uni-mainz.de}
\emailAdd{anna.pollmann@uni-wuppertal.de}
\emailAdd{mrongen@uni-mainz.de}

\abstract{The Wavelength-shifting Optical Module (WOM) is a novel optical sensor that uses wavelength shifting and light guiding to substantially enhance the photosensitive area of UV optical modules. It has been designed for the IceCube Upgrade, a seven-string extension of the IceCube detector planned for the 2022/2023 South Pole deployment season. The WOM consists of a hollow quartz cylinder coated in wavelength shifting paint which serves as detection area and has two photomultipliers (PMTs) attached to the end faces. The light-collecting tube increases the effective photocathode area of the PMTs without producing additional dark current, making it suitable for low-signal, low-noise applications. 
We report on the design and performance of the WOM with a focus on the 12 modules in production for deployment in the IceCube Upgrade.
While the WOM will be deployed in IceCube, its design is applicable to any large-volume particle detector based on the detection of Cherenkov light.

\vspace{4mm}
{\bfseries Corresponding authors:}
John Rack-Helleis$^{1*}$, Anna Pollmann$^{2}$, Martin Rongen$^{1}$\\
{$^{1}$ \itshape Institute for Physics, JGU Mainz, D-55122 Mainz, Germany}\\
{$^{2}$ \itshape Department of Physics,Bergische Universität Wuppertal, Gaußstraße 20, Wuppertal, Germany}\\[4mm]
$^*$ Presenter
}

\FullConference{$37^{\mathrm{th}}$ International Cosmic Ray Conference (ICRC 2021)\\
July 12th -23rd, 2021\\
Online - Berlin, Germany}

\begin{document}
\maketitle

\section{The IceCube Neutrino Observatory and the IceCube Upgrade}
\label{ICU}

The IceCube Neutrino Observatory is a cubic-kilometer scale neutrino detector installed in the ice at the geographic South Pole \cite{IceCube}. Reconstruction of direction, energy, and flavor of the neutrinos relies on the optical detection of Cherenkov radiation emitted by charged particles produced in the interactions of neutrinos in the surrounding ice or the nearby bedrock.
    
The IceCube Upgrade \cite{IcecubeUpgrade} is planned to be deployed during the 2022/2023 South Pole Season and marks the first extension of the IceCube detector since its completion in 2010. Over 700 additional modules - the majority of which use the  mDOM \cite{mDOM2021} and the D-Egg \cite{dEgg2021} designs - will be deployed on seven additional strings (chains of modules in one drill hole). The spacing between optical modules in the Upgrade will be $\sim$\SI{20}{\m} horizontally and \SI{3}{\m} vertically, compared to $\sim100$~m horizontally and 17~m vertically in IceCube. This denser instrumentation lowers the energy threshold for neutrino detection, improves event reconstruction, and allows for a more precise calibration of the detector medium (the ice). Additionally, it will be used for in-situ studies of novel optical sensor designs.
Among these will be the Wavelength-shifting optical module (WOM) of which we are preparing to deploy 12 units.

\section{The IceCube Upgrade WOM}
\label{WOM}

In order to instrument large detector volumes with photo-sensitive sensors capable to resolve single photons, the general approach is to use photomultiplier tubes in the detector modules.
This means that the financial costs as well as the background noise rate scale linearly with the deployed sensitive area. By shifting the detection area from PMTs to a photon-capturing tube instead, the instrumented area is increased significantly by elongating the tube.
The light captured by the tube can be focused on small photocathode areas, lowering the thermionic noise contribution in comparison to  designs using PMTs only. 

\begin{figure}[h]
\centering
\includegraphics[width=0.7\textwidth]{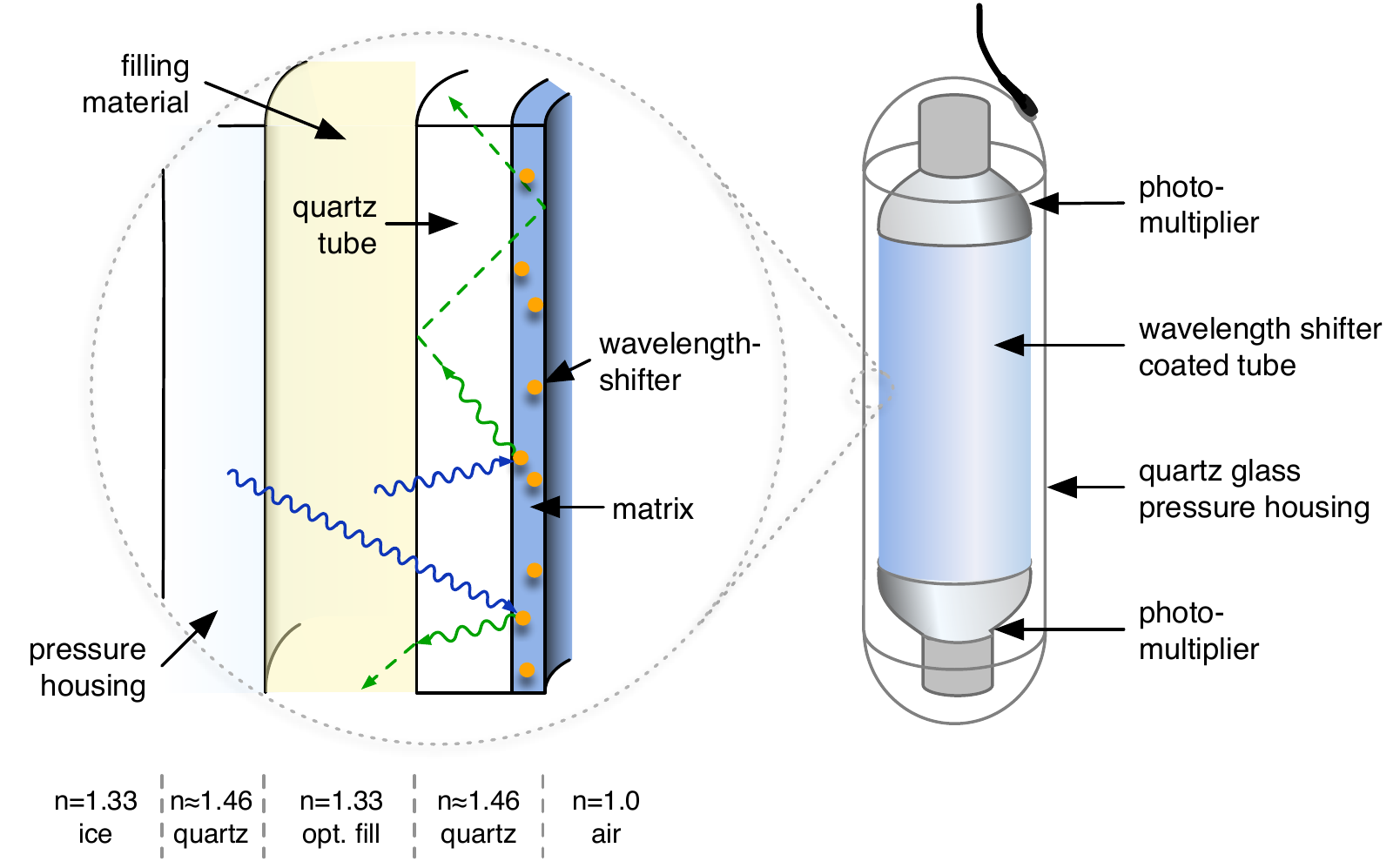}
\caption{Sketch and working principle of the WOM. UV Photons are absorbed, shifted and reemitted by the wavelength shifting substrate. Reemitted photons are guided to the read-out PMTs by means of total internal reflection.}
\label{fig:WOM_sketch}
\end{figure}

The WOM, schematically shown in \autoref{fig:WOM_sketch}, consists of a transparent tube which is coated with paint containing wavelength-shifting (WLS) organic luminophores. UV photons incident on the tube are absorbed in the paint layer with high efficiency and re-emitted isotropically at optical wavelengths.
\begin{wrapfigure}{l}{0.5\textwidth}
    \centering
    \includegraphics[width=0.5\textwidth]{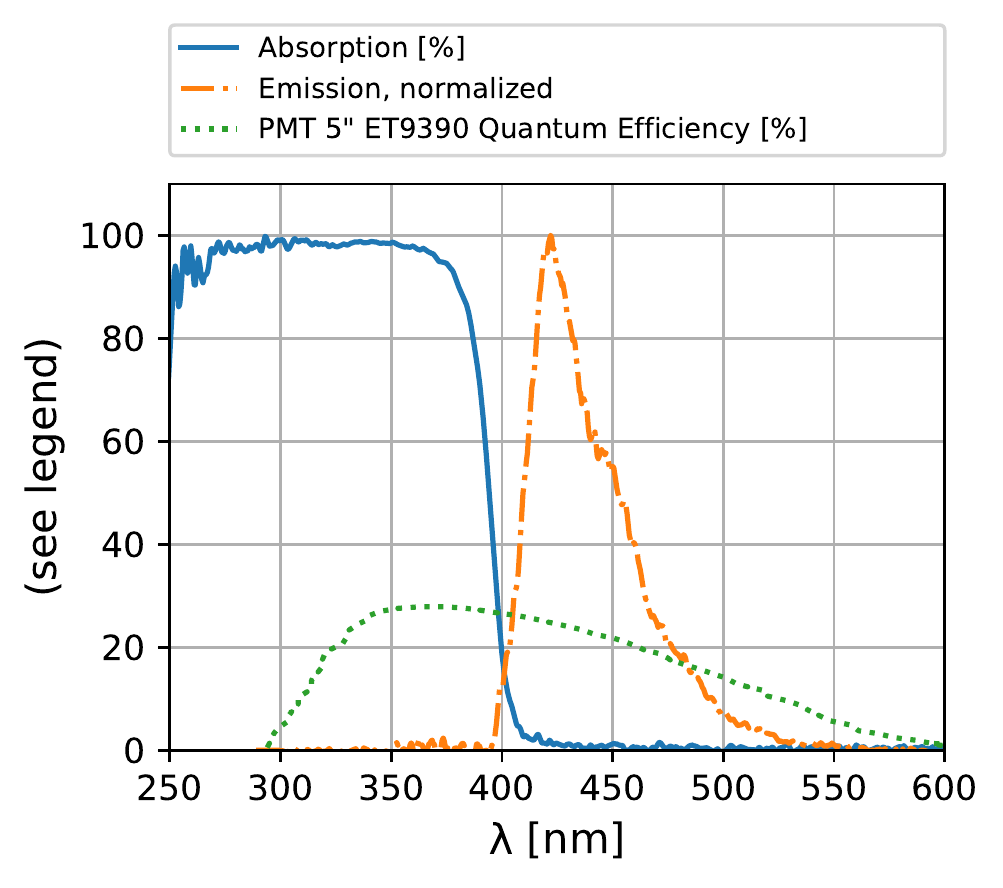}
    \caption{Absorption and emission profile of the WLS-paint in blue and orange. The quantum efficiency of the 5" ET9390 is shown in green.}
    \label{fig:WLSspectrum}
\end{wrapfigure}
Absorption and emission spectrum of the paint are shown in relation to the PMT quantum efficiency in \autoref{fig:WLSspectrum}.
If the photon emission angle is above the critical angle for total internal reflection, the photon is trapped in the tube and guided towards the read-out PMTs at the end faces.
In comparison to the mDOM and the D-Egg, the WOM offers two unique features: low overall noise rates, as the photocathode area is small in comparison to the sensitive area of the WLS tube, and enhanced UV sensitivity. The latter is especially desirable in IceCube, since the Cherenkov spectrum peaks in the UV range.  

The WOM concept was first introduced in \cite{Schulte:2013dza}, the wavelength-shifting paint and the coating process have been thoroughly investigated in \cite{WLS-paper,VLVNT15}, with testing procedures outlined in \cite{Peiffer:2017vsm}. Here we discuss the specific developments required for the design and production of IceCube Upgrade WOMs. The choice of PMT and DAQ are discussed in \autoref{PMT}. The performance of a prototpye WLS tube is discussed in \autoref{WLS}.
For the IceCube Upgrade, the WOM and the required  electronics need to be assembled within a UV-transparent quartz housing serving as pressure vessel. This requirement as well as the mandatory filling material between the pressure housing and WLS tube are the content of \autoref{pressure_housing} and \autoref{filling_material}. Assembly of all components is described in \autoref{assembly}.

\section{Photomultiplier and data acquisition}
\label{PMT}

Since the diameter of the PMT constrains the diameters of the wavelength shifting tube as well as the pressure housing, the choice of PMT model is a critical design decision. The following requirements were considered:
\begin{itemize}
    \setlength\itemsep{-0.5em}
    \item  A reasonably large diameter to make best use of the available drill hole diameter
    \item  Good efficiency at the edge of the photocathode, needed for coupling to the WLS tube
    \item  Low thermionic noise to demonstrate good noise characteristics for supernova detection
    \item  Gain $>$\SI{5e6}{} at safe voltages, for single photon detection using the selected DAQ
    \item Flat cathode surface to ease gluing of the tube
\end{itemize}

Given these requirements, the \emph{Electron Tubes ET9330} \footnote{\url{https://et-enterprises.com/images/data_sheets/9390B.pdf}} was identified as the closest match and is currently under detailed investigation.

In order to enable easy integration into the IceCube Upgrade computing and communication infrastructure,
the analog signals from the PMTs will be read out using \emph{waveform micro-bases} \cite{LOM2021}. An IceCube \emph{Mini-Mainboard} is used to handle communication on the string. The high voltage for the PMT is supplied using a staged Cockroft-Walton chain implemented on the \emph{waveform micro-base}  and as already used on the \emph{micro-bases} for the mDOM 3" PMTs\cite{mDOM2021}. 

\begin{figure}
    \centering
    \begin{minipage}{0.49\textwidth}
        \centering
        \includegraphics[width=\textwidth]{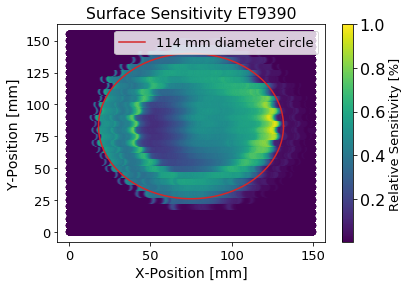} 
        \caption{Relative surface sensitivity of the ET9390 Photocathode at 900V supply voltage.}
        \label{fig:PMT_surface}
    \end{minipage}\hfill
    \begin{minipage}{0.49\textwidth}
        \centering
        \includegraphics[width=0.95\textwidth]{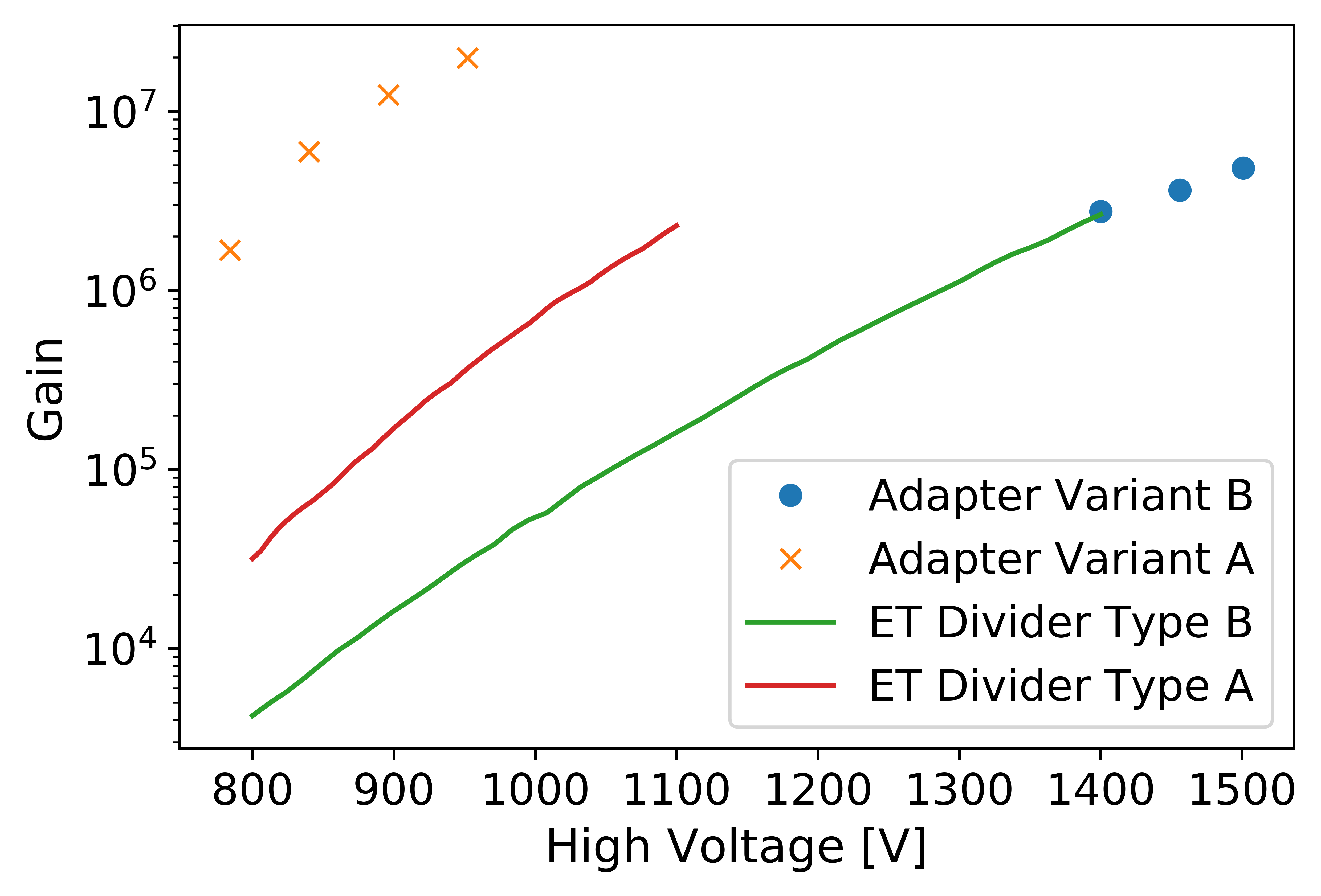} 
        \caption{Gain measurement (dots and crosses) and data sheet values (lines) for different divider configurations.}
        \label{fig:gains_PMT}
    \end{minipage}
\end{figure}

Custom adapter printed circuit board (PCBs) were developed to accommodate for the differences in pin-out and required voltage ratios. Following the manufacturer's recommendations, a low gain variant was successfully implemented (see \autoref{fig:gains_PMT}, labeled Adapter Variant B).
Implementing the desired high gain configuration (similar to ET Divider Type A) has so far not been successful as the resulting single-photon charge distributions are very broad with no discernible valley. 

\begin{figure}
    \centering
    \begin{minipage}{0.49\textwidth}
        \centering
        \includegraphics[width=\textwidth]{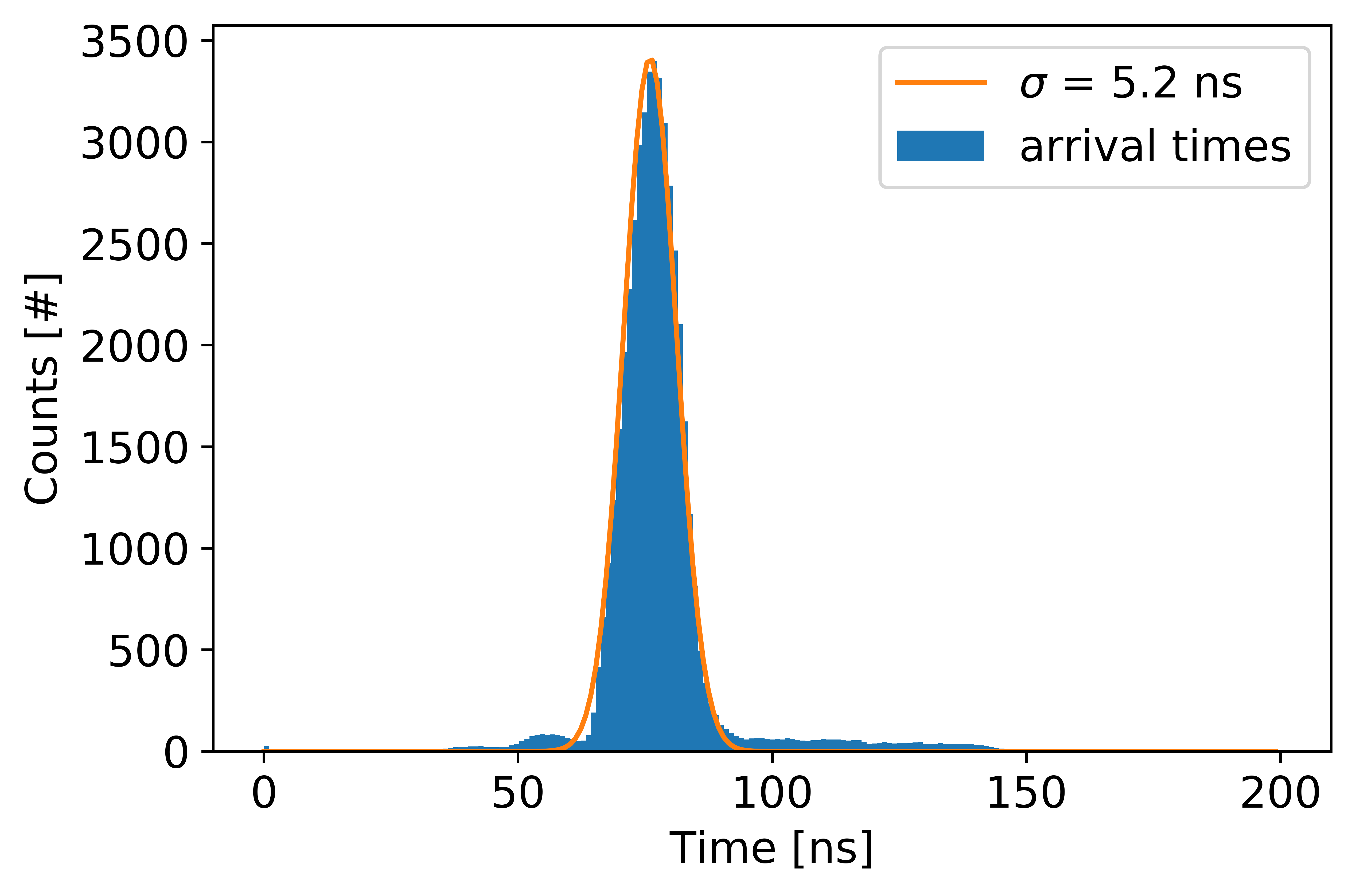}
        \caption{Response time spread of the Upgrade WOM PMT (ET9390).}
        \label{fig:timing_PMT}
    \end{minipage}\hfill
    \begin{minipage}{0.49\textwidth}
          \centering
             \includegraphics[width=\textwidth]{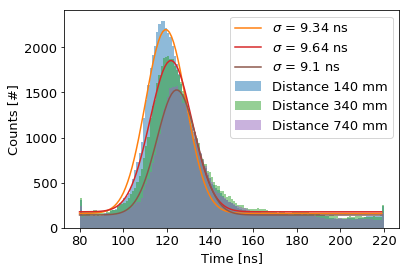}
            \caption{Photon arrival time distribution for selected z coordinates. The WLS tube has a length of \SI{76}{\cm}}
            \label{fig:WOM_timing}
    \end{minipage}
\end{figure}

Using the voltage ratios as directly available from the \emph{micro-base} (3:1:1:1:1, see adapter variant A in \autoref{fig:gains_PMT}) results in a high gain of $10^7$ at a bias of \SI{\approx 1000}{\V}, well below the recommended operating voltage. The resulting small potential between photocathode and first dynode is unfavorable as it is likely to result in an inhomogeneous collection efficiency and strong magnetic field dependencies.
To gauge this effect, the surface sensitivity was measured using an attenuated pulsed laser mounted to a translation stage. The relative sensitivity map is shown in \autoref{fig:PMT_surface}. The photoactive area is approximately confined to a \SI{114}{\mm} diameter circle. The asymmetry in the inner diameter is attributed to the expected inhomogeneous collection efficiency caused by the low bias voltage.
\autoref{fig:timing_PMT} shows the PMT transit time spread using the adapter variant A. The standard deviation is \SI{5.2}{\ns} and thus smaller than the spread introduced in the WLS tube (see \autoref{timing}). Further properties still to be studied include the PMT's thermionic noise as a function of temperature as well as dynamic range.

\section{Wavelength-shifting Tube}
\label{WLS}

The WLS-tube consists of quartz glass\footnote{\url{https://www.heraeus.com/media/media/hca/doc_hca/products_and_solutions_8/solids/Solids_HSQ300_330MF_EN.pdf}} tube which is \SI{76}{\cm} long and \SI{10.6}{\cm} in outer diameter. The two 5" ET9330 PMTs and the WLS-tube are optically coupled using a UV curing glue\footnote{\url{https://www.norlandprod.com/adhesives/NOA146H.html}}.
 
 \subsection{Efficiency}
\begin{figure}
    \centering
    \begin{minipage}{0.49\textwidth}
        \includegraphics[width=0.9\textwidth]{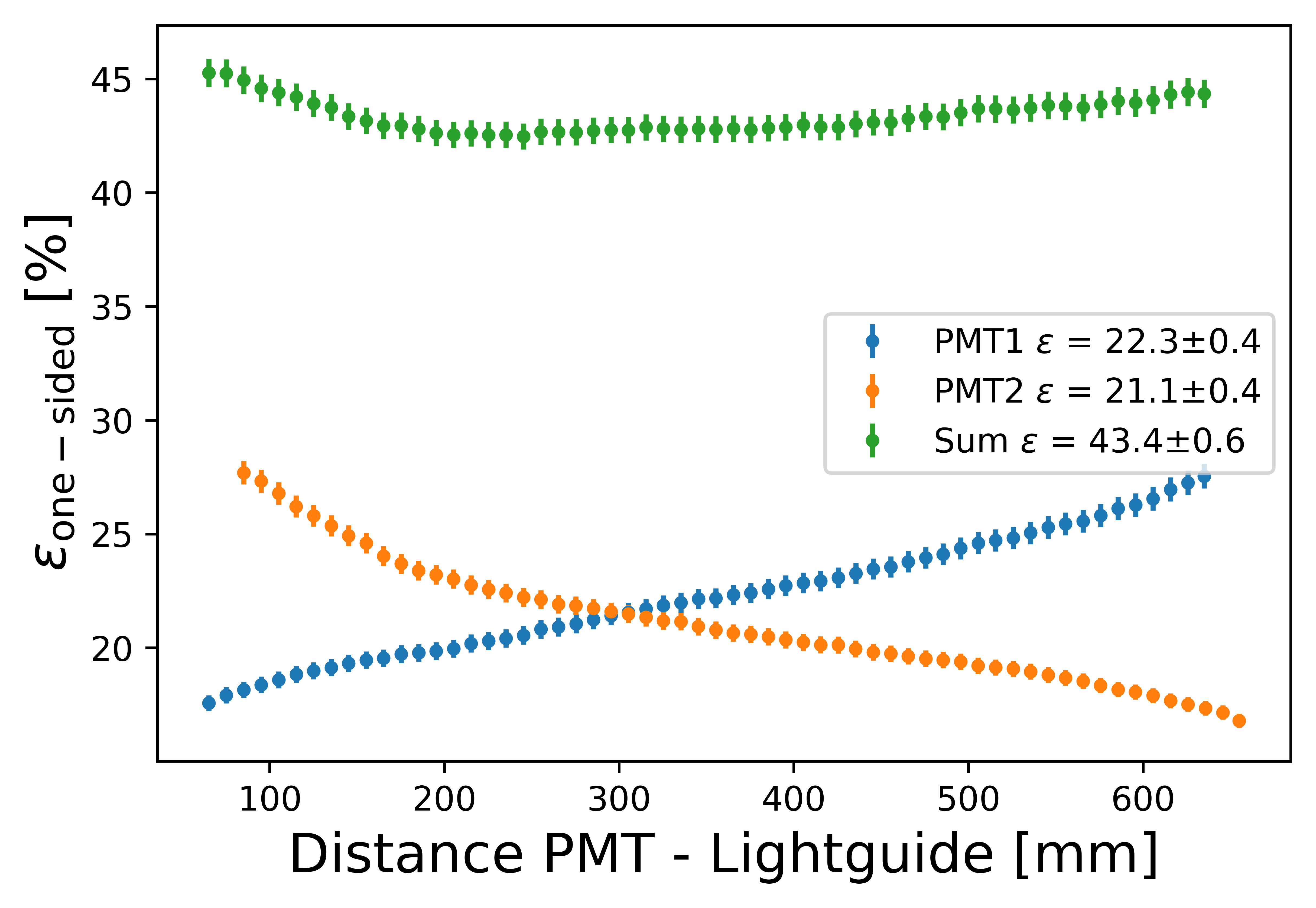}
        \caption{Light guiding efficiency of the inner Tube as a function of the $z$-coordinate. The orange and blue curves represent the individual efficiencies at the PMTs and green denotes the overall guiding efficiency.}
        \label{fig:linear_scan}
  \end{minipage}\hfill
  \begin{minipage}{0.49\textwidth}
        \centering
        \includegraphics[width=0.9\textwidth]{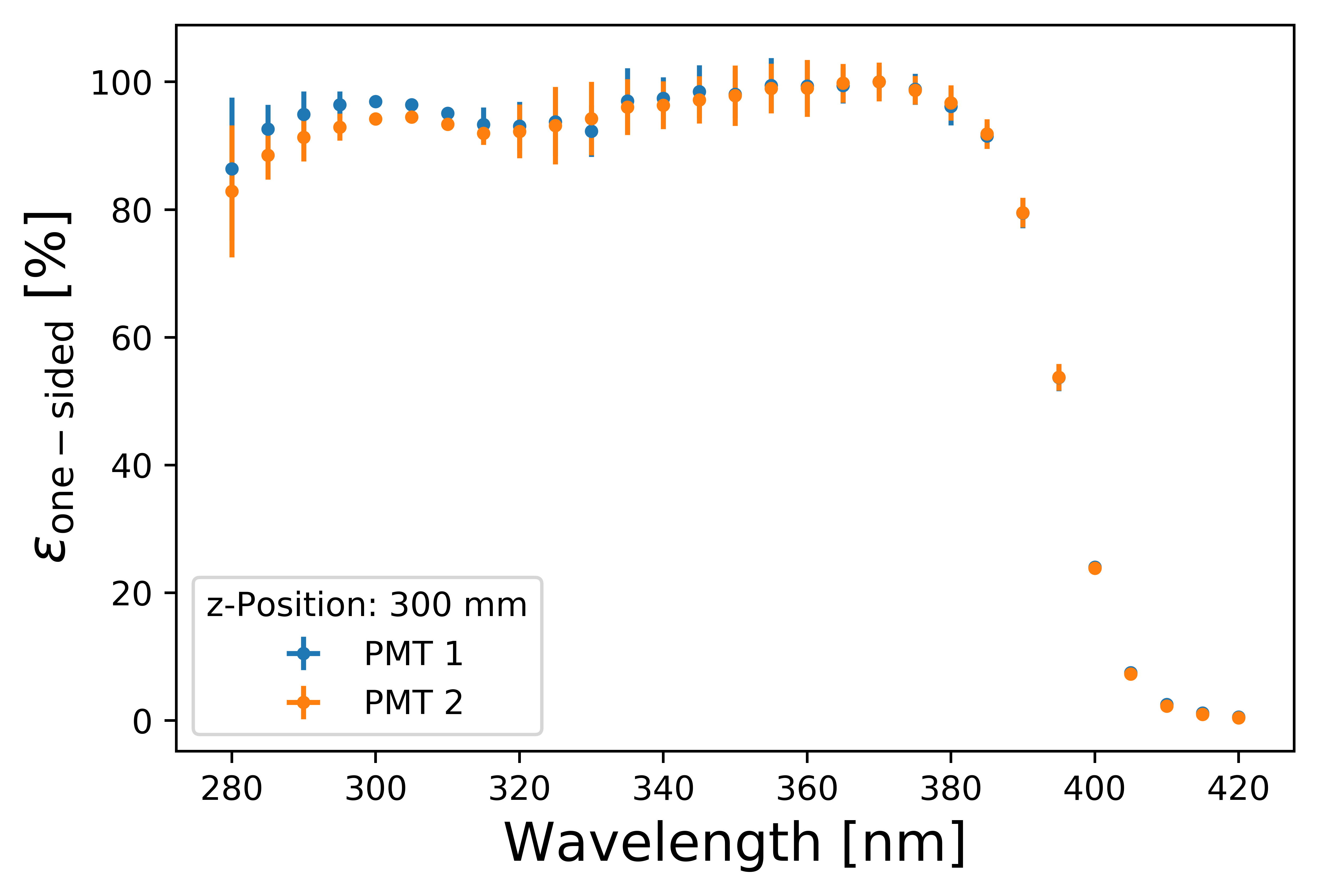}
        \caption{Capture efficiency of the WLS tube as a function of the incident light's wavelength. The measurement was performed in the middle of the tube and is corrected for transport losses.}
        \label{fig:wavelength_scan} 
    \end{minipage}
\end{figure}

 To determine the WLS-tube efficiency as a function of distance along the tube (z-coordinate) and wavelength, a setup similar to the one described in \cite{Peiffer:2017vsm} is used:
 The output of a xenon arc lamp is wavelength selected using a monochromator. The beam is chopped and its intensity is controlled using a photodiode read out by lock-in amplifiers. Using a movable liquid light guide allows for measuring the WOM's efficiency as a function of $z$ and $\lambda$. The efficiency dependence on the $z$-coordinate is shown in \autoref{fig:linear_scan}. It can be concluded that approximately \SI{40}{\percent} of the photons re-emitted by the WLS paint reach the readout PMTs.
 The measurement in \autoref{fig:wavelength_scan} shows that the process of absorption and re-emission of the wavelength-shifting paint has an efficiency of close to \SI{100}{\percent} in the wavelength range from \SIrange{280}{400}{\nm}.  

\begin{figure}
  \centering
      \begin{minipage}{0.49\textwidth}
        \centering
        \includegraphics[width=0.9\textwidth]{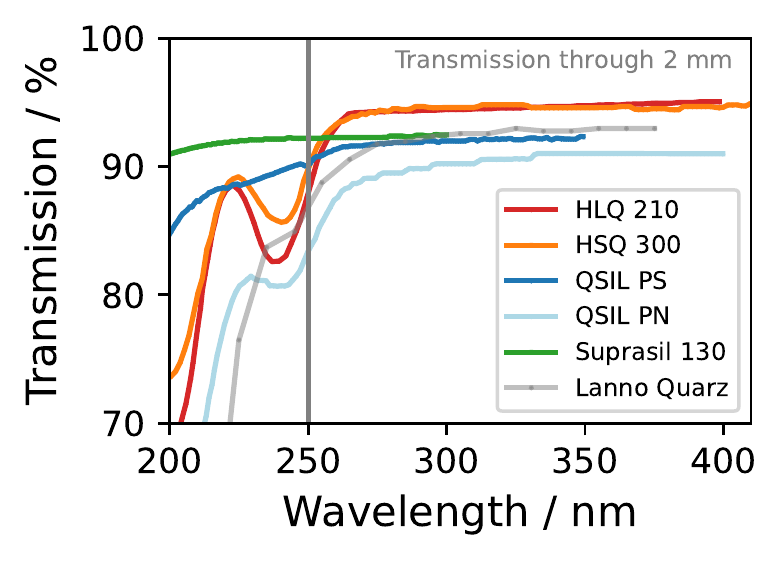} 
        \caption{Measurement of the transmission of several glass samples considered for the pressure vessel. The interesting region lies above \SI{250}{\nm} ( grey line).}
        \label{fig:transmission_housing}
    \end{minipage}\hfill
  \begin{minipage}{0.49\textwidth}
        \includegraphics[width=0.9\textwidth]{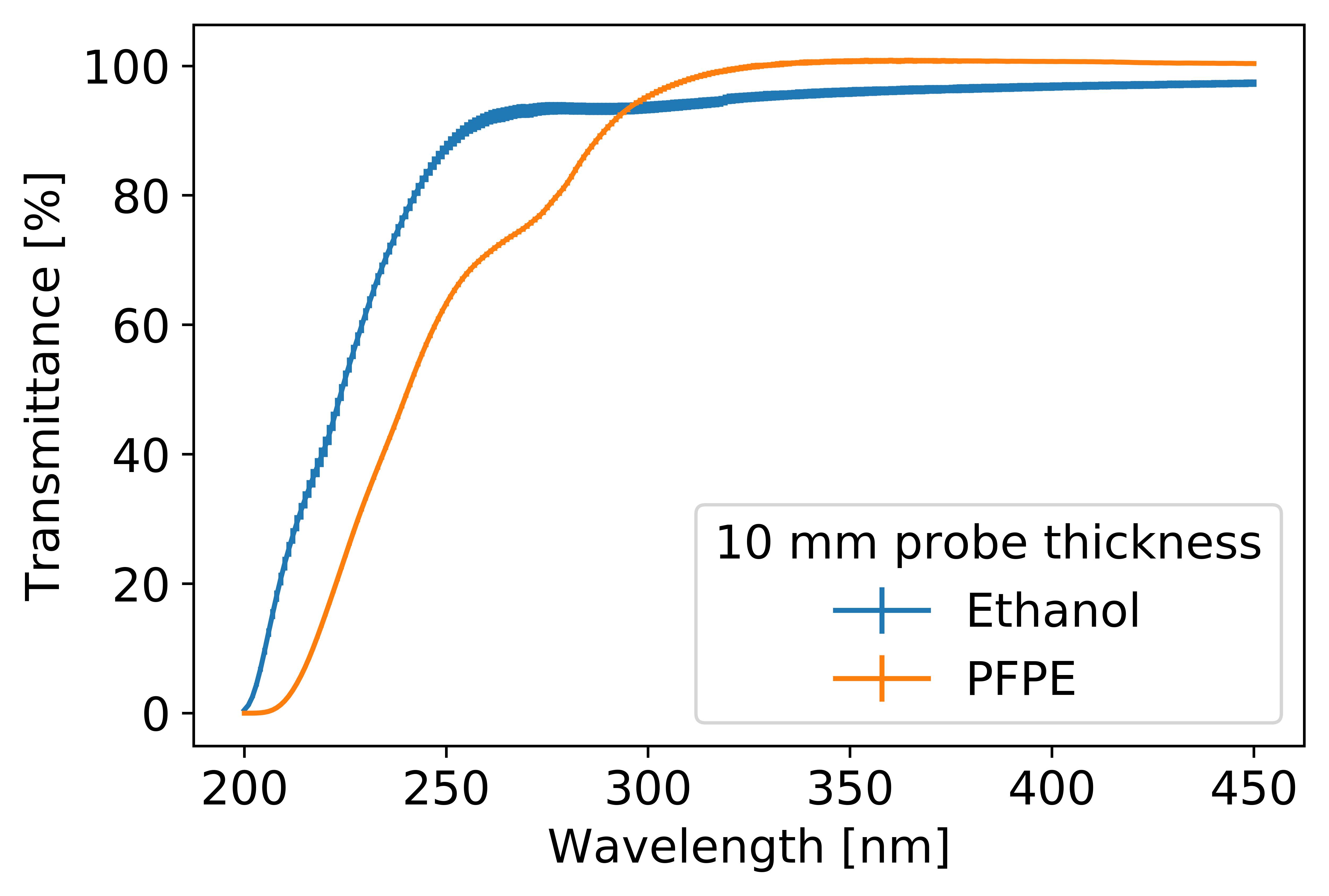} 
        \caption{Transmission of selected filling materials for the upgrade WOM.}
        \label{fig:transmission_filling}
    \end{minipage}
\end{figure}

 \subsection{Timing}
\label{timing}

The enhanced UV-sensitivity and low noise of the WOM come at the cost of a $2\times$ wider distribution in photon arrival times compared to a bare PMT. Here, the major contribution is the travel time of the photons inside the WLS tube prior to reaching the PMTs. The absorption and re-emission of the WLS paint plays a minor role due to its fast decay time of \SI{1.35}{\ns} \cite{WLS-paper}.
The arrival time distribution of the WLS has been measured using a pico-second pulser\cite{pulser} as light source and a standalone ADC to digitize the PMT signals. In \autoref{fig:WOM_timing}  the arrival time distribution for selected distances between the light entry point and readout PMT is shown. It can be observed, that for larger distances, the average arrival time shifts to later times, matching the time shift due to the effective speed of light in quartz glass. The overall timing resolution is approximated to be \SI{10}{\ns}.

\section{Pressure Housing}
\label{pressure_housing}

The choice of the pressure housing is critical, since low UV transmittance or background from the glass vessel due to radioactive impurities would significantly deteriorate the device's performance. In earlier prototypes, the isotope $^{238}\mathrm{U}$ has been proven to cause significant background contribution.

Glass samples from several companies were investigated. The transparency was measured using a calibrated test stand and samples of different thickness. The results can be seen in \autoref{fig:transmission_housing}. The transmission above $250\,\text{nm}$ exceeds 90\% for almost all samples, rendering all shown options viable. For cost effectiveness reasons, the glass HLQ210 was chosen as material for the vessel.

Neutron activation was used to estimate the radioactive backgrounds in the glass to be approximately \SI{6}{\becquerel\per\kg} amounting to a dark noise rate of \SI{114}{\Hz} for a \SI{\sim19}{\kg} pressure vessel. This contribution is negligible compared to the lowest background expected from the PMTs\footnote{\url{https://et-enterprises.com/images/data_sheets/9390B.pdf}}.


\section{Filling material}
\label{filling_material}

The optimal filling material between the pressure vessel glass and the inner WLS tube was chosen based on two derivations: the effective area of the device was first calculated analytically accounting for all Fresnel transmissions as well as the total internal reflections in the WLS tube assuming homogeneous illumination with plane waves. Second, a raytracing Monte Carlo simulation was modeled taking these effects into account. 

\begin{wrapfigure}{l}{0.5\textwidth}
    \centering
    \includegraphics[width=0.5\textwidth]{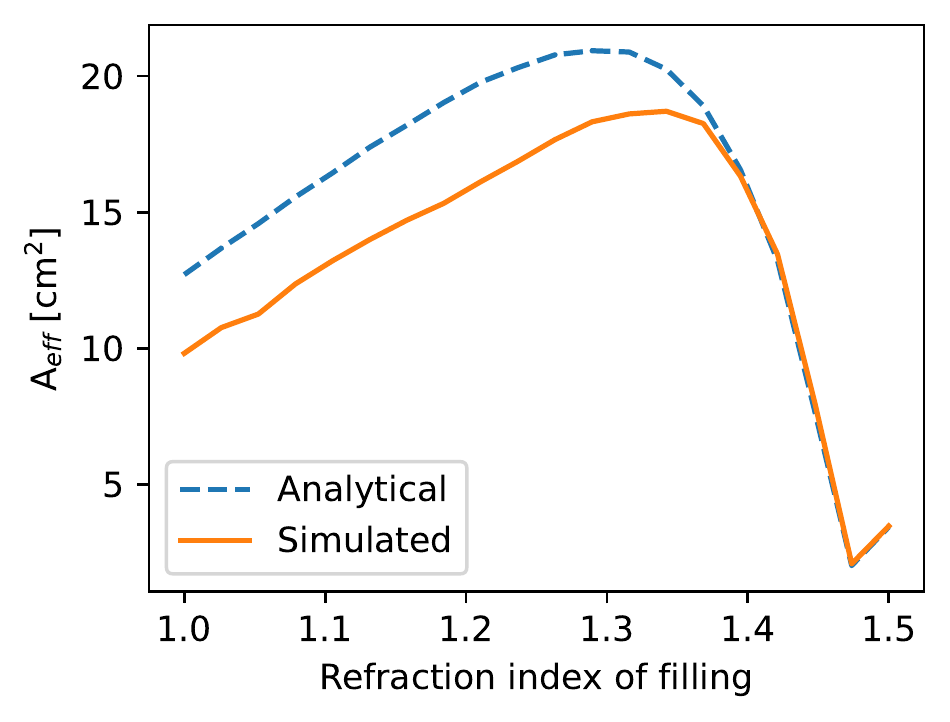} 
    \caption{Effective area of the fully module as a function of the refractive index $n$ of the filling material obtained in two ways: an analytical derivation (blue) as well as a MC simulation of the device (orange).}
    \label{fig:Aeff_WOM}    
\end{wrapfigure}

\autoref{fig:Aeff_WOM} shows the change in effective area as a function of the refractive index of the filling material. While the absolute difference between the approaches is still under investigation,
they both demonstrate the optimum refractive index of the filling material is $n=1.33$. This opens up several choices for the filling material with Ethanol and PFPE (Perfluorpolyether) being the most prominent. We prefer PFPE over Ethanol due to the higher vapor pressure and chemical inertness. Its refractive index is $n = 1.30$ and it shows good transparency in the UV, as shown in \autoref{fig:transmission_filling}.   

\section{WOM Assembly}
\label{assembly}

\begin{figure}[h]
\centering
\includegraphics[origin=c,width=15cm,keepaspectratio]{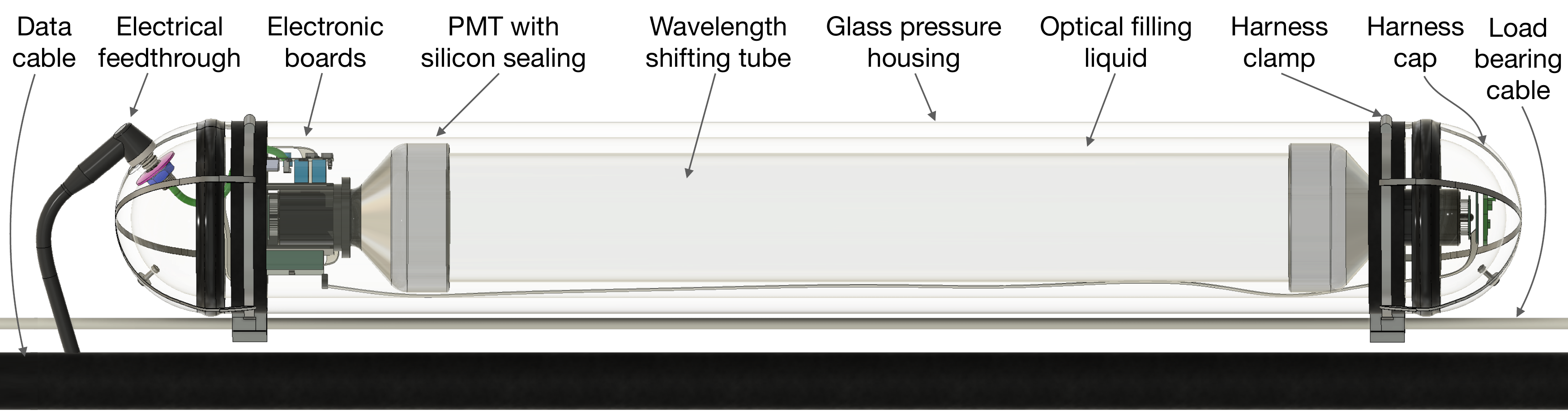}
\caption{CAD drawing of the assembled WOM module.}
\label{fig:WOM_CAD}
\end{figure}

The general mechanical design for the IceCube Upgrade WOM is shown in \autoref{fig:WOM_CAD}. The outer assembly features a quartz\footnote{\url{https://www.heraeus.com/media/media/hca/doc_hca/products_and_solutions_8/solids/Solids_HSQ300_330MF_EN.pdf}} pressure vessel together with borosilicate endcaps which are attached to the pressure vessel by pulling a vacuum of approximately \SI{0.5}{\bar} on the inside. A penetrator on the top end cap allows for communication and powering of the device.

The module has to withstand a pressure of up to $700\,\text{bar}$ during deployment in the ice. 
To ease handling and enable pressure testing, the length of the glass vessel was restricted to $1.3\,\text{m}$. This in turn limits the available space for electronics as well as the length of the inner tube. The outer diameter of the vessel is $173\,\text{mm}$, set by the 5" PMTs. 
The vessel is fixed on the string by two pipe clamps connecting to custom clamps on a load-bearing cable. One cable clamp acts as guide only, to allow the cable to extend 1\% in length under load. In order to prevent vertical movement of the module, which slightly shrinks under pressure, metal bands are crossed above the end caps and welded to the pipe clamps. 

The weight of the full module is mainly due to the approximately $19\,\text{kg}$ of the vessel.  
The inner part consists of the WLS tube with attached PMTs, as described in \autoref{WLS}. The two PMTs are embedded in a silicone\footnote{\url{https://www.smooth-on.com/tb/files/MOLD_STAR_15_16_30_TB.pdf}} wrapper which matches the inner diameter of the pressure housing. The silicone confines the filling material\footnote{\url{https://www.solvay.com/en/product/galden-ht-170}} to the space between the PMTs. Feedthrough holes for electronic cables as well as a pressure valve allow for pouring of the filling material and for pressure compensation during deployment.\\

\section{Summary and Outlook}
\label{Aeff}
The Wavelength-shifting optical module is a UV-sensitive, low-noise photosensor which is being developed for the IceCube Upgrade. With this design, we obtain an effective area of approximately $19\,\text{cm}^2$. It is most sensitive to photons between \SIrange{280}{400}{\nm}. 
The aim is to deploy the devices in clusters within the IceCube Upgrade array to locally lower the energy threshold, thus enhancing the resolution of low energy neutrino reconstructions. The reduced average noise also improves the sensitivity to supernova neutrinos, which are not individually resolvable but result in a small temporal rate increase over the rate of background events.
The large and easily scalable effective area of the WOM in combination with the low inherent noise makes it interesting for other applications, e.g. instrumenting large veto volumes for particle detectors. The WOM is already incorporated into the design of the ShiP \cite{shipProposal} experiment, based on a modified version without pressure vessel and using SiPMs instead of PMTs \cite{SHip_2019}.

\bibliographystyle{ICRC}
\bibliography{thebibliography}
\clearpage
\section*{Full Author List: IceCube Collaboration}




\scriptsize
\noindent
R. Abbasi$^{17}$,
M. Ackermann$^{59}$,
J. Adams$^{18}$,
J. A. Aguilar$^{12}$,
M. Ahlers$^{22}$,
M. Ahrens$^{50}$,
C. Alispach$^{28}$,
A. A. Alves Jr.$^{31}$,
N. M. Amin$^{42}$,
R. An$^{14}$
% ...
\\

\noindent
% ...
$^{12}$ Universit{\'e} Libre de Bruxelles, Science Faculty CP230, B-1050 Brussels, Belgium \\
% ...
$^{14}$ Department of Physics and Laboratory for Particle Physics and Cosmology, Harvard University, Cambridge, MA 02138, USA \\
% ...
$^{17}$ Department of Physics, Loyola University Chicago, Chicago, IL 60660, USA \\
$^{18}$ Dept. of Physics and Astronomy, University of Canterbury, Private Bag 4800, Christchurch, New Zealand \\
% ...
$^{22}$ Niels Bohr Institute, University of Copenhagen, DK-2100 Copenhagen, Denmark \\
% ...
$^{28}$ D{\'e}partement de physique nucl{\'e}aire et corpusculaire, Universit{\'e} de Gen{\`e}ve, CH-1211 Gen{\`e}ve, Switzerland \\
% ...
$^{31}$ Karlsruhe Institute of Technology, Institute for Astroparticle Physics, D-76021 Karlsruhe, Germany  \\
% ...
$^{42}$ Bartol Research Institute and Dept. of Physics and Astronomy, University of Delaware, Newark, DE 19716, USA \\
% ...
$^{50}$ Oskar Klein Centre and Dept. of Physics, Stockholm University, SE-10691 Stockholm, Sweden \\
% ...
$^{59}$ DESY, D-15738 Zeuthen, Germany \\
% ...

\subsection*{Acknowledgements}

\noindent
USA {\textendash} U.S. National Science Foundation-Office of Polar Programs,
U.S. National Science Foundation-Physics Division,
U.S. National Science Foundation-EPSCoR,
Wisconsin Alumni Research Foundation,
Center for High Throughput Computing (CHTC) at the University of Wisconsin{\textendash}Madison,
Open Science Grid (OSG),
Extreme Science and Engineering Discovery Environment (XSEDE),
Frontera computing project at the Texas Advanced Computing Center,
U.S. Department of Energy-National Energy Research Scientific Computing Center,
Particle astrophysics research computing center at the University of Maryland,
Institute for Cyber-Enabled Research at Michigan State University,
and Astroparticle physics computational facility at Marquette University;
Belgium {\textendash} Funds for Scientific Research (FRS-FNRS and FWO),
FWO Odysseus and Big Science programmes,
and Belgian Federal Science Policy Office (Belspo);
Germany {\textendash} Bundesministerium f{\"u}r Bildung und Forschung (BMBF),
Deutsche Forschungsgemeinschaft (DFG),
Helmholtz Alliance for Astroparticle Physics (HAP),
Initiative and Networking Fund of the Helmholtz Association,
Deutsches Elektronen Synchrotron (DESY),
and High Performance Computing cluster of the RWTH Aachen;
Sweden {\textendash} Swedish Research Council,
Swedish Polar Research Secretariat,
Swedish National Infrastructure for Computing (SNIC),
and Knut and Alice Wallenberg Foundation;
Australia {\textendash} Australian Research Council;
Canada {\textendash} Natural Sciences and Engineering Research Council of Canada,
Calcul Qu{\'e}bec, Compute Ontario, Canada Foundation for Innovation, WestGrid, and Compute Canada;
Denmark {\textendash} Villum Fonden and Carlsberg Foundation;
New Zealand {\textendash} Marsden Fund;
Japan {\textendash} Japan Society for Promotion of Science (JSPS)
and Institute for Global Prominent Research (IGPR) of Chiba University;
Korea {\textendash} National Research Foundation of Korea (NRF);
Switzerland {\textendash} Swiss National Science Foundation (SNSF);
United Kingdom {\textendash} Department of Physics, University of Oxford.

\end{document}